# Jet absorption and corona effect at RHIC. Extracting collision geometry from experimental data

V. S. Pantuev

*University at Stony Brook, Stony Brook, NY 11794-3800, USA*

*on leave from Institute for Nuclear Research, Russian Academy of Sciences, Moscow, Russia*

We demonstrate a possible existence of a finite formation time of strongly interacting plasma in nuclear collisions at RHIC from recent experimental data. To show this, we use a simple model based on Monte Carlo simulation of nucleus-nucleus collisions with realistic nuclear density distribution. The most striking feature of the experimental data - an absence of absorption of high transverse momentum pions in the reaction plane direction for mid-peripheral collisions - points to the presence of a surface zone with no absorption and strong suppression in the inner core. A natural interpretation of such a zone could be the plasma formation time $T \simeq$ 2-3 fm/c. The existence of a formation time could dramatically change our understanding of many experimentally observed features. With this assumption we describe the angular anisotropy of high transverse momentum pions with respect to the reaction plane and the centrality dependence of nuclear modification factor in Au+Au and Cu+Cu collisions.

PACS: 25.75.Nq, 52.27.Gr

Nowadays, the physics of relativistic nuclear collisions at RHIC goes from the phase of discoveries [1] to the stage of understanding properties of a new state of nuclear matter: the quark-gluon plasma, QGP. Naive expectations that QGP to be a gas of weakly interacting quark and gluons was washed out by very first experimental data. Collective phenomena, such as radial and elliptic flow, reveal "hydro" properties of the QGP, which behave "like a good liquid rather than a dilute gas of quasi particles" [2] and the idea was introduced that plasma at RHIC should be in strongly coupled regime, sQGP. Another indication that sQGP behave more like a liquid could be found in experimental data on modification of the away side jet shape, which becomes much wider and possibly might be developed by a sonic-boom in a shape of Mach cone [3]. One of the predicted phenomenon to be seen at RHIC is jet quenching [4]: hard partons were expected to lose some energy in the formed medium. Very first RHIC data have indeed shown significant particle suppression at high $p_T$ [1]. In addition, in contrast to early expectations for weakly interacting QGP, the data indicate strong coupling of heavy quarks, charm and bottom, to the medium [5].

Recently, the PHENIX collaboration published another intriguing result of $R_{AA}$ dependence on the azimuthal angle $\phi$ relative to the reaction plane in Au+Au collisions at $\sqrt{s_{NN}}$=200 GeV [6], Fig. 1. $R_{AA}$ is a nuclear modification factor, which is defined as a number of the observed jets normalized to the expected number of jets from the superposition of individual nucleon-nucleon collisions:

$$R_{AA}(p_T) = \frac{(1/N_{evt})\, d^2 N^{A+A}/dp_T d\eta}{(\langle N_{binary}\rangle / \sigma_{inel}^{N+N})\, d^2 \sigma^{N+N}/dp_T d\eta}, \quad (1)$$

where $\langle N_{binary} \rangle$ is the average number of binary nucleon-nucleon collisions in a particular centrality class.

The most interesting feature of these data is that, at event centrality class 50-60%, for transverse momenta above 4 GeV/c, in-plane $R_{AA}$ equals to one within the errors. This implies no absorption at all for high $p_T$ pions. For such high momenta, Cronin effect [7] is negligible and can not bring $R_{AA}$ close to one. At the same time, a significant particle absorption is seen in out-of-plane. At this event centrality class the amount of nuclear matter is still significant in all directions. It's puzzling that a high momentum parton can "punch through" the interaction zone in the in-plane direction but may be stopped in the other direction. Apparently, parton energy loss calculations can't describe this feature of in-plane $R_{AA}$=1, e.g., see Fig.5 in [8]. In this paper we offer a possible explanation of these and certain other features of the data.

We consider a simple model using the Monte Carlo simulation of nucleus-nucleus collisions based on the Glauber approach with a Woods-Saxon nuclear density distribution and discussed in detail in [9]. We restrict ourselves to the data of high $p_T$ pions with transverse momentum above 4 GeV/c, where $R_{AA}$ does not depend on $p_T$. We assume that all high $p_T$ pions are produced

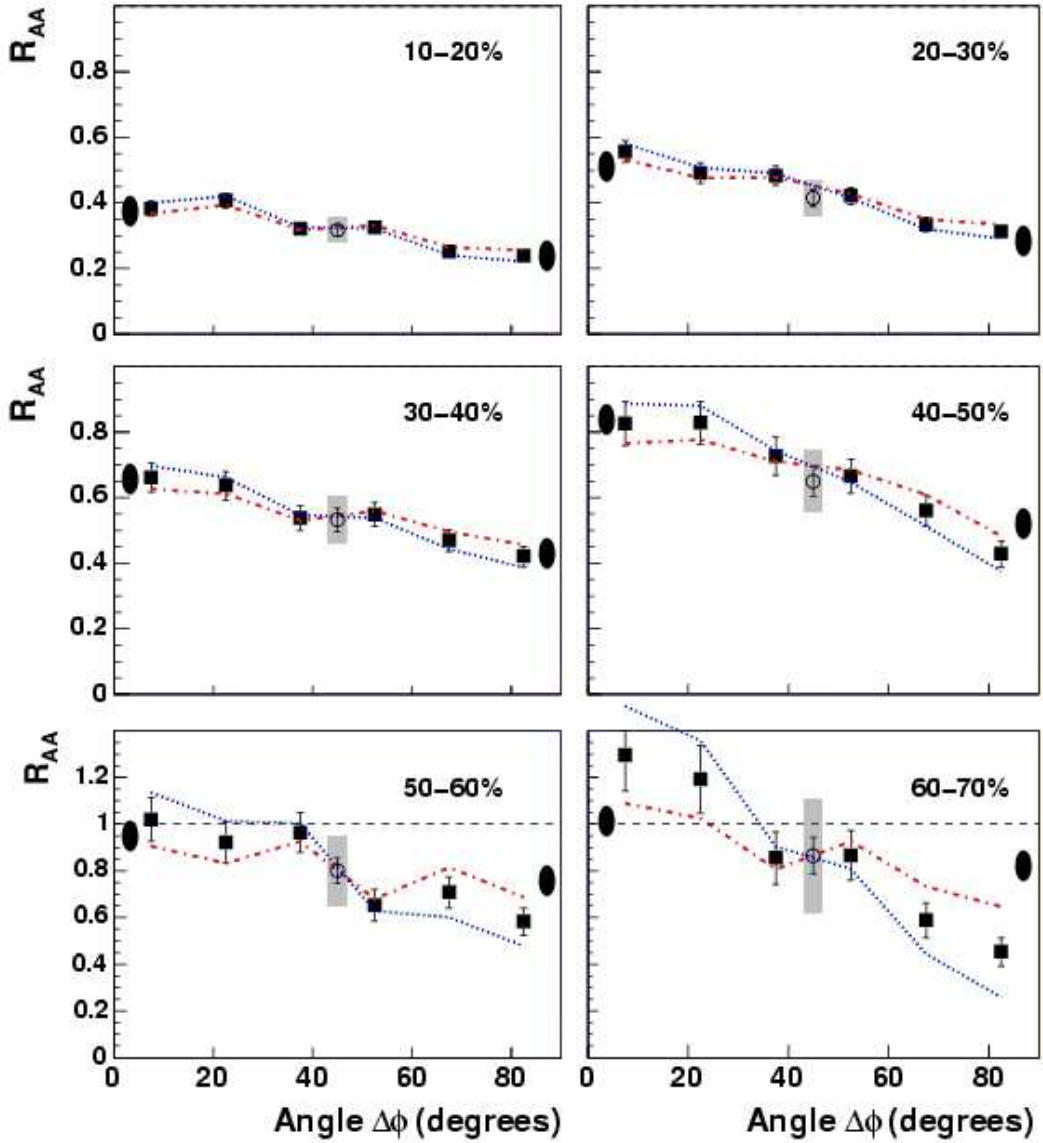

Fig.1 PHENIX $R_{AA}$ results for $\pi^0$ at momenta 5-8 GeV/c versus angle $\phi$ relative to the reaction plane for different centralities, from more central 10-20% to peripheral 60-70% Au+Au collisions at RHIC [6]. Points are experimental data, thin lines show systematic errors from the reaction plain resolution, vertical bars in the middle show averaged over the reaction plane $R_{AA}$ value and it's error. Black ovals are predictions of our model

by parton fragmentation and that the number of hard partons is proportional to $N_{binary}$. If there is no absorption, $R_{AA}$=1 in all directions and the shape of the event is isotropic. To explain the experimentally observed feature of in-plane $R_{AA}$=1, we investigate the role of purely geometrical factors. In our model, jets which have to travel through the medium at *some direction* from their production point to the surface less than a distance $L$ will leave the interaction zone unmodified. Jets originating in the core region deeper than $L$ suffer significant energy loss and are completely absorbed. The whole picture looks like a pure *corona* jet production, but we allow this corona region to be larger than a Woods-Saxon type skin. The cut-off parameter $L$ should be of the order of the size of the in-plane interaction zone at 50-55% centrality, about 2-3 fm.

The nuclear modification factor versus angle $\phi$, $R_{AA}(\phi)$, by definition is a single particle inclusive parameter. This is a measure of the number of partons $dN$ produced in a particular direction within $d\phi$. To estimate this number, we do the following: first, we generate a spatial distribution of parton production points

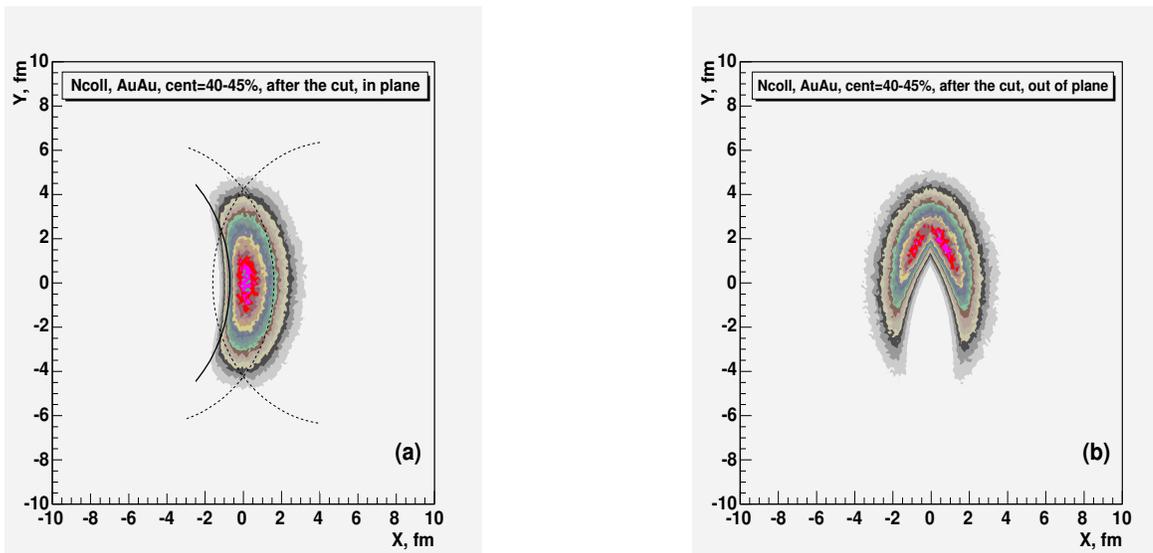

Fig.2. Distribution of collision vertexes, $N_{coll}$, in the transverse plane for the Au+Au 40-45% centrality class for the surviving partons. The impact parameter vector is oriented along the x axis; the beam direction is perpendicular to the page. (a) for partons, produced in-plane (horizontally, $\phi=0$) *and* to the observer direction positioned to the right side from the collision. Dashed lines show the envelope of the Woods-Saxon nuclear radii, solid line is our cut. (b) partons are moving vertically ($\phi=\pi/2$) *up* to the observer, who looks at the event out-of-plane *and* in the direction to the top of the plot

by Monte Carlo simulation of nucleon-nucleon binary collisions, $N_{coll}$, transverse to the beam direction plane. This forms an "almond" shaped interaction region. The time evolution of this shape during the first 2-3 fm/c is less than 20%, see Fig.7 in [10] and can be neglected. Second, we have to select the direction toward the observer and its orientation with respect to the reaction plane for a particular collision event. For example, in Fig. 2 (a), the observer positioned exactly in the reaction plane on the right side. We draw a cut edge through this "almond" at depth $L$ from the observer side. In Fig. 2 (a) we apply such a cut by stepping to the left by distance $L$ from the envelope of the Woods-Saxon nuclear radius. All jets, produced to the right of this cut, traveling toward the observer will escape without any interaction. All partons produced in the *same* direction but deeper than $L$ will be completely absorbed by the medium and are not plotted in the figure. The observer can't see part of the back side of the collision. A mirrored picture will be seen by the observer from the left. We can calculate $R_{AA}$ as a ratio of the *seen* collisions, $N_{coll}$, to the total $N_{binary}$.

A more complicated production zone appears when one tries to detect particles out of the reaction plane direction, as shown in Fig. 2 (b). This is the case when the observer detects particles exactly out of plane in the direction to the top of the plot. Here the cut edge goes through the "almond" at distance $L$ down from the envelope of the Woods-Saxon radii. A significant portion of the collisions in the central region is cut off. Consequently, the number of produced high $p_T$ particles $out-of-plane$ will be smaller than those $in-plane$. We estimate $R_{AA}$ as an average of the $R_{AA}^{in}$ for $in-plane$ and $R_{AA}^{out}$ for the $out-of-plane$ case.

Jet energy loss can not happen in an infinitely thin layer. Therefore, we smooth the cut edge (say, in Fig. 2 (a), it is the left edge). An arbitrary weight function in the form of a Fermi distribution with diffuseness parameter $a$ was applied: $weight(l) = \frac{1}{1+exp((L-l)/a)}$. The parameter $a$ was varied from 0.01 to 0.5 fm. The results do not change significantly in this parameter range. A maximum deviation of 5% was obtained for $a$=0.5 fm, only in very peripheral collisions. The default value was chosen to be $a$=0.2 fm, which produces a diffuseness of the cut edge at about 1 fm. There is one free parameter $L$ in our calculation, which was adjusted to get $R_{AA}^{in}$=0.9±0.1 for 50-55% centrality, yielding $L$=2.3±0.6 fm. This is close to what is seen experimentally and leaves some room for Cronin enhancement [7], if any.

Particle distribution in the azimuthal direction can be described by the amplitude $v_2$ of the second Fourier coefficient, dN/d$\phi$=N(1+2$v_2$cos(2$\phi$)). All other components are known to be small [11]. *A priori*, within the frame of our model, there may not be an exact $cos(2\phi)$ dependence. Thus, we estimate the value of $v_2$ as $v_2=1/4(R_{AA}^{in} - R_{AA}^{out})/R_{AA}$. In other words, $v_2$ is determined by the jet survived probability in and out of plane. By an additional investigation, we found that in our model $R_{AA}$ has almost perfect $cos(2\phi)$ shape.

The results of our calculations are shown in Fig. 1 and Fig. 3. We can successfully describe the data for $R_{AA}$ in- and out-of-plane for all centrality classes. The parameter $v_2$ reaches 11-12% in mid-central events (centrality 30-35%) and nicely follows the trends observed in the experiments [12, 13]. Our result disproves the assertion that jet quenching models can not explain the measured $v_2$ at high $p_T$ [14].

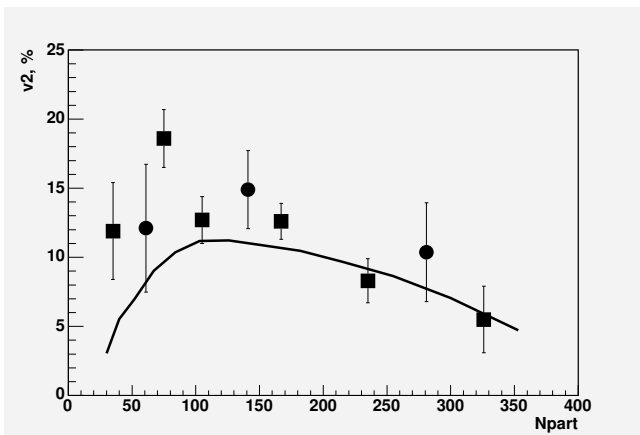

Fig.3. Calculated ellipticity parameter $v_2$ for Au+Au collisions, solid line, versus the number of participant nucleons, $N_{part}$. Data for $\pi^0$ with error bars are: circles for 4.5-9 GeV/c, squares for 5-7 GeV/c. PHENIX preliminary data [12, 13]

We investigated the sensitivity of our result to various assumptions. First, we consider the thickness of the material integrated over the path length ($\propto \rho dl$) as a critical cut-off parameter. The centrality dependence becomes very strong in this case and can not describe the data. We find similar disagreement using a quadratic dependence of the absorption cut ($\propto \rho l dl$). Another premise tested uses the number of participant nucleons, $N_{part}$, instead of $N_{coll}$. In this case the centrality dependence is weaker than in the experimental data with a maximum value of $v_2$ of 5%.

In Fig. 4, we plot our results for $R_{AA}$ at high $p_T$ in Cu+Cu collisions at 200 GeV using the same $L=2.3$ fm. It is worth mentioning that calculations were done before preliminary experimental Cu+Cu data were presented [16]. Fig. 4 shows our estimate for Au+Au and Cu+Cu collisions in comparison with PHENIX experimental data.

Within our model we can also describe $R_{AA}$=0.36 for neutral pions in Au+Au collisions at $\sqrt{s_{NN}}$=62.4 GeV [8] in most central events as measured by PHENIX. In this case we get even larger value $L$=3.5 fm and $v_2$ reaches 11%.

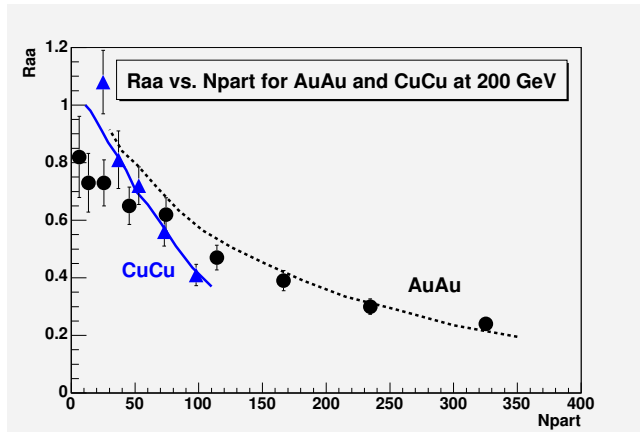

Fig.4. $R_{AA}$ for Au+Au (dashed curve) and Cu+Cu (solid curve) collisons versus the number of participant nucleons, $N_{part}$. The cicles are experimental $\pi^0$ data for Au+Au collisions integrated for $p_T \geq 4$ GeV/c [15]. The triangles are data for Cu+Cu collisions of $p_T \geq 7$ GeV/c [16]. Only statistical errors are shown

What could be the physical interpretation of the geometrical cutoff $L$? Our guess is that it is not actually spatial, but a *time* cutoff, $T = L/c$. If this were the parton formation time, it should be momentum dependent, but in the experiment $R_{AA}$ is essentially constant for momenta above 3-4 GeV/c. It is more natural to assign this parameter $T$ to a "plasma" formation time, or, at least, the time when parton energy loss actually starts. This gives a simple and elegant interpretation of the effect: particles produced close to the surface of the collision zone have time, about 2-3 fm/c, to escape. After that time a very dense and strongly interacting matter is formed and this matter absorbs high $p_T$ partons. To prove that this is indeed a time, we performed another test calculation. Let all partons fly in all possible transverse directions with speed of light for a time $T$=2.3 fm/c. Some partons will leave interaction zone, some will go in to this zone. In our scenario all partons, which still remain inside interaction zone, determined by Woods-Saxon radii envelop, after that time should be absorbed. Performing such calculation, we found exactly the same numbers which are presented in Fig. 4.

The time we found is almost an order of magnitude larger than many of the theoretical models used, especially hydrodynamic models [10]. It is worth to mention that time parameter $\tau$ in such models is applied locally, where hydrodynamics starts to develop *local* pressure or equilibrium, thus meaning of this parameter $\tau$ is different.

An interesting explanation was proposed by E. Shuryak and J. Liao [17]. They suggest that jet energy loss is small "until the matter cools down" to form a liquid or even color "polymer chains", which as any phase transition needs some latent time.

Undistorted production of high $p_T$ particles from the corona region explains the lack of change of hadron distributions in forward jets with centrality and orientation relative to the reaction plane [18]. The appearance of large $v_2$ for high $p_T$ mesons simply from the collision geometry means that the contribution of elliptic flow to $v_2$ should diminishes at $p_T$ above 4-5 GeV/c.

In conclusion, we present an experimentally based alternative to the paradigm of traditional parton energy loss models. Based on PHENIX experimental data for reaction plane angular dependence of $R_{AA}$, we have estimated the thickness of the "corona" in high $p_T$ particle production. We described the $R_{AA}$ and the $v_2$ centrality dependence well for Au+Au collisions and made predictions for the Cu+Cu case. The large $v_2$ seen in experiments at high $p_T$ can be assigned to the collision geometry and strong absorption, not flow of high $p_T$ particles. We attribute the visible "corona" thickness to the plasma formation time.

This work was partially supported by the US-DOE grant DE-FG02-96ER40988. The author would like to thank Barbara Jacak for useful discussions and Jiangyong Jia, who generated histograms of the Monte Carlo simulation. These histograms form the base of our calculations.